\documentclass[11pt,a4paper]{article}
\usepackage{jcappub,bm}
\usepackage{booktabs}

\title{
Next-to-leading resummation of cosmological perturbations via the Lagrangian picture: 2-loop correction in real and redshift spaces}
\author[a]{Tomohiro Okamura,} \author[b,c]{Atsushi Taruya}
\author[d,e]{and Takahiko Matsubara}

\affiliation[a]{Astronomical Institute, Tohoku University, Aoba
  Aramaki Aoba, Sendai, 980-8578, Japan}

\affiliation[b]{ Research Center for the Early Universe, School of
  Science, University of Tokyo, Bunkyo-ku, Tokyo 113-0033, Japan}

\affiliation[c]{ Institute for the Physics and Mathematics of the
  Universe, University of Tokyo, Kashiwa, Chiba 277-8568, Japan}

\affiliation[d]{Kobayashi-Maskawa Institute for the Origin of
  Particles and the Universe, Nagoya University,
  Chikusa, Nagoya, 464-8602, Japan}
\affiliation[e]{ Department of Physics, Nagoya University,
  Chikusa Nagoya, 464-8602, Japan}

\emailAdd{t-okamura@astr.tohoku.ac.jp}
\abstract{ We present an improved prediction of the
  nonlinear perturbation theory (PT) via the Lagrangian picture, which
  was originally proposed by Matsubara (2008). Based on
  the relations between the power spectrum in standard PT and that in
  Lagrangian PT, we derive analytic expressions for the
  power spectrum in Lagrangian PT up to 2-loop order in both real
  and redshift spaces. Comparing the improved prediction of Lagrangian 
    PT with $N$-body simulations in real space, 
  we find that the 2-loop corrections can extend the
  valid range of wave numbers where we can predict the power
  spectrum within $1$ \% accuracy by a factor of 1.0 ($z=0.5$), 1.3 ($1$),
  1.6 ($2$) and 1.8 ($3$) vied with 1-loop Lagrangian PT results. On the
  other hand, in all redshift ranges, the higher-order corrections 
  are shown to be less significant on the two-point correlation functions 
  around the baryon acoustic peak, because the 1-loop Lagrangian 
  PT is already accurate enough to explain the nonlinearity on those 
  scales in $N$-body simulations.}


\arxivnumber{1105.1491}

\begin{document}
\maketitle

%==================================================%
\section{Introduction}
\label{sec:intro}
%==================================================%

  In the forthcoming era of precision cosmology, 
  accurate and precise predictions of cosmological theories play
  crucial roles not only to estimate the cosmological parameters but 
  also to detect a tiny signal from the  observational data. 
  The large-scale structure of the universe is a
  powerful probe of cosmology, and recently the baryon acoustic
  oscillations (BAOs) imprinted on the large-scale structure
  \cite{EHT98,mat04,eis05} have attracted much attention 
  to constrain the nature of dark energy. 
  
  There are many ongoing and planned redshift surveys aiming at 
    clarifying the nature of the dark energy through a precision 
    measurement of BAOs, including the Sloan Digital Sky
  Baryon Oscillation Spectroscopic Survey (BOSS)
  \cite{BOSS}, Hobby-Eberly Telescope Dark Energy Experiment (HETDEX)
  \cite{HETDEX}, BigBOSS \cite{BigBOSS}, Subaru Measurement of Images
  and Redshifts (SuMIRe) \cite{SuMIRe}, Wide-Field Infrared Survey
  Telescope (WFIRST) \cite{WFIRST}, Euclid \cite{Euclid}.
  To measure the characteristic scales of BAO precisely enough
  for investigating the dark energy, accurate
  theoretical template for the power spectrum and correlation function 
  on BAO scales is quite essential.

  Although the nonlinear gravitational clustering of galaxies on BAO scales is 
  rather moderate, the linear
  theory of density perturbations \cite{pee80} is inadequate
  to quantitatively describe the BAOs \cite{MWP99,SE05}.
  The standard perturbation theory (SPT)
  provides a systematic way to investigate the gravitational clustering of dark
  matter in nonlinear regime
  \cite{jus81,vis83,fry84,GGRW86,SM91,MSS91,JB94,SF96,BCGS02}. Even
  though the SPT describes the nonlinearity fairly well at
  sufficiently high redshifts \cite{JK06}, it is still 
  insufficient to describe the clustering on BAO scales at
  observationally relevant redshifts, $z=0$--$3$. To overcome such
  situation, various methods to improve the SPT are proposed,
  partially resumming infinite series of higher-order perturbations. 
  The renormalized perturbation theory (RPT) \cite{CS06a,CS06b,CS08} 
  is a representative pioneering work, and since it appeared, various
  approaches have been subsequently proposed 
  \cite{mcd07,val07,MP07,MP08,TH08,pie08,mat08a,TH09}. The quantitative 
  predictions of those approaches are compared with $N$-body simulations, 
  and actually shown to excellently reproduce the $N$-body results quite
  well in a certain range of power spectrum and two-point correlation 
  function. The accuracy and precision of the predictions actually depend
  on the treatment of each method, and a comprehensive test of various 
  methods is examined in ref.~\cite{Carlson09}.

  Most of the above newly developed perturbation theories predict the
  nonlinear gravitational clustering of dark matter in real space. However, the
  observable quantity in actual redshift surveys is the clustering of
  galaxies in redshift space. There are two ingredients to be added to
  the perturbation theory in order to compare the theoretical predictions
  with observations: redshift-space distortions \cite{kai87,Ham98,S04} 
  and biasing \cite{DEFW85,BBKS}. 
  While the redshift distortions induce the clustering anisotropies, 
  the galaxy biasing can produce a scale-dependent enhancement (or suppression) 
  in the clustering amplitude. Hence, coupled with the effect of 
  gravitational clustering, both the amplitude and shape of the 
  power spectrum and two-point correlation function would be significantly 
  changed, and an accurate modeling of BAOs that ensures a percent-level
  precision seems non-trivial. 

  The resummed PT via the Lagrangian picture provides a unique framework
  to include both the ingredients into the fundamental formulation, and
  does not have any phenomenological parameter of dynamics once the 
  biasing model is fixed~\cite{mat08a,mat08b,mat11} (but see \cite{taruya10}). 
    In contrast to the SPT formulated as Eulerian PT, 
    the resummed PT by refs.~\cite{mat08a,mat08b,mat11} is based on the 
    framework of the Lagrangian perturbation theory (LPT)
  \cite{buc89,mou92,buc92,BE93,buc94,HBCJ95,cat95,CT96,EB97}, 
  in which the displacements of fluid element
  are treated as dynamical variables. It is known that the 
  first-order LPT reproduces the classic Zel'dovich approximation 
  \cite{zel70}. In both Eulerian and Lagrangian PTs, 
  the small variables in the perturbative expansion
  correspond to the initial density fluctuations. Therefore, when the
  theoretical predictions of LPT are made in the Eulerian space with the
  full expansion of perturbations, LPT should give mathematically 
  equivalent results to the SPT expansion \cite{mat11}. 
  In the Lagrangian resummation method, correlations of the displacement
 field at zero-lag are kept in the exponent, and other contributions to
 the power spectrum are expanded as usual in Eulerian space.
    This partial expansion corresponds to the resummation of a 
    class of infinite higher-order perturbations. As a result, 
    an exponential prefactor naturally arises in the expression of the 
    nonlinear power spectrum \cite{mat08a}.  The 
  exponential prefactor plays an important role in describing the
  nonlinear smearing effects of BAO in both real and redshift spaces.

  In previous works, the Lagrangian resummation method has been investigated only
  up to 1-loop order in perturbations. The purpose of this paper is to
  generalize the previous work and to carry out 2-loop calculations of
  the same resummation method. We derive analytic expressions of the
  2-loop results in both real and redshift spaces. 
  To see how well the 2-loop corrections improve the previous 1-loop 
    results, we focus on the power spectrum of dark matter in 
    real space, and compare the predictions with numerical
    simulations. Although we leave a quantitative analysis 
  in redshift space for future work, 
  the improvement of the predictions shown in this paper is an important 
  step, and would lead to a strong motivation to actually evaluate the 
  effects of redshift-space distortions and the biasing 
  via the LPT up to 2-loop order.

  This paper is organized as follows. In section~\ref{sec:prep}, useful
  equations and concepts, which are mostly derived in previous work,
  are summarized. In section~\ref{sec:analytic}, our main results of the
  2-loop calculations both in real and redshift spaces are presented.
  In section~\ref{sec:compare}, the analytic expressions are compared with
  $N$-body simulations of dark matter. In section~\ref{sec:summary}, our
  results and conclusions are summarized.

%==================================================%
\section{Lagrangian Perturbation Theory and Standard Perturbation
  Theory \label{sec:prep}
}
%==================================================%

  In this section, we summarize some concepts and equations of Lagrangian 
  resummations method, which we use in this paper. Readers find derivations
  of the following equations in refs.~\cite{mat08a,mat08b}.

  According to the result of 1-loop resummation via LPT, 
  the following identity is
  confirmed by straightforward calculations of 1-loop SPT and LPT: 
%EEEEEEEEEEEEEEEEEEEEEEEEEEEEEEEEEEEEEEEEEEEE%
\begin{multline}
P_{\rm LPT}(k)
=\exp{\left[-\frac{k^{2}}{6\pi^{2}}\int dp\,P_{\rm L}(p) \right]} 
\times\left[ P_{\rm L}(k)+P_{\rm{SPT}}^{1\mbox{\scriptsize
        -loop}}(k)+\frac{k^{2}}{6\pi^{2}}P_{\rm L}(k)\int dp\,P_{\rm
      L}(p)\right],
\label{eq:onelooplpt}
\end{multline}
%EEEEEEEEEEEEEEEEEEEEEEEEEEEEEEEEEEEEEEEEEEEE%
  where $P_{\rm LPT}(k)$ is the nonlinear power spectrum predicted
  by the 1-loop resummation via LPT, $P_{\rm L}(k)$ is the linear
  power spectrum, and $P_{\rm SPT}^{1\mbox{\scriptsize -loop}}(k)$ is the
  1-loop contributions (without linear contribution) to the power
  spectrum predicted by the SPT. Note that 
  the time dependence of the power spectrum $P_{\rm LPT}$ is implicitly 
  encoded in the linearly extrapolated power spectrum $P_{\rm L}(k)$  
  evaluated at the redshift of our interest.

  As we mentioned in section~\ref{sec:intro}, 
  the displacement field is the fundamental variable in LPT, and is treated 
  as a small perturbative quantity. In computing the power spectrum, 
  we need to evaluate the correlation of displacement fields at 
  different positions in the exponent. The resummation method proposed by 
  refs.~\cite{mat08a,mat08b} gives a specific but efficient recipe to 
  treat this. That is, the correlation of the displacement fields at a 
  single position is kept in the exponent, while the correlation 
  among components of separated displacement vectors, which is expected to 
  be small compared to zero-lag correlation as long as we consider 
  a large separation, is expanded from the exponent. 
  The exponential prefactor in eq.~\eqref{eq:onelooplpt} indeed  
  represents the zero-lag correlation up to the 1-loop level. 
  It is easily checked that expanding the exponential prefactor and 
  truncating the series up to the 1-loop order exactly recovers 
  the power spectrum of 1-loop SPT.

  Note that in the language of RPT~\cite{CS06a,CS06b}, the resummation 
  treatment by refs.~\cite{mat08a,mat08b} corresponds to a partial vertex 
  renormalization of the SPT expansion~\cite{mat08a}, and  
  the exponential prefactor in eq.~\eqref{eq:onelooplpt} 
  contains the information beyond the 1-loop SPT.   
  Further, the resultant functional form of the exponential prefactor 
  quite resembles the renormalized propagator 
  at high-$k$ limit \cite{CS06b}, and is shown to play a physically 
  important role in describing the nonlinear smearing effect of BAOs.  
  Nevertheless, because of a rather strong damping, the 
  resultant predictions for the 
  power spectrum ceases to work well at higher-$k$ modes, where the 
  non-linear enhancement of the power spectrum amplitude 
  becomes significant. Thus, similar to RPT and other improved PT 
  treatment, the accuracy of the Lagrangian resummation 
  treatment will be limited to a specific range  $k\lesssim k_{\rm damp}$, 
  where $k_{\rm damp}$ is the characteristic scale of the exponential damping 
  factor. But, this characteristic scale depends on the order of loop 
  corrections included in the power spectrum expression, and including the 
  next-to-leading order will definitely improve the predictions applicable 
  to some higher-$k$ modes.

 The main purpose of this paper is to extend the 1-loop expression 
    \eqref{eq:onelooplpt} to a general relation between the power spectrum 
    in SPT and that in LPT for arbitrary loop-order, and is to 
    explicitly derive the 2-loop expression. To do so, we first notice that 
  the exponential prefactor corresponds to the square of a Fourier
  transform of the one-point probability distribution function of the 
  displacement field, $\langle
  e^{-i\bm{k}\cdot\bm{\varPsi}}\rangle^2$, where $\bm{\varPsi}(\bm{q})
  = \bm{x}(\bm{q})-\bm{q}$ is the 
  displacement field, $\bm{x}$ and $\bm{q}$
  are respectively Eulerian and Lagrangian coordinates of a fluid
  element \cite{mat08b,mat11}. According to the cumulant expansion
  theorem \cite{ma85}, this factor reduces to
%EEEEEEEEEEEEEEEEEEEEEEEEEEEEEEEEEEEEEEEEEEEE%
\begin{equation}
  \left\langle e^{-i\bm{k}\cdot\bm{\varPsi}} \right\rangle
  = \exp\left[-\sum_{n=1}^\infty\frac{k_{i_1}\cdots k_{i_{2n}}}{(2n)!}
      A_{i_1\cdots i_{2n}}^{(2n)} \right],
\label{eq:FourierPDF}
\end{equation}
%EEEEEEEEEEEEEEEEEEEEEEEEEEEEEEEEEEEEEEEEEEEE%
where
%EEEEEEEEEEEEEEEEEEEEEEEEEEEEEEEEEEEEEEEEEEEE%
\begin{align}
A^{(2n)}_{i_{1}\cdots i_{2n}}= \left[\prod_{i=1}^{2n}\int \frac{d^{3}p_{i}}{(2\pi)^{3}}\right] \delta^{3}_{\rm D}\left(\sum_{i=1}^{2n}\bm{p}_{i}\right)C_{i_{1}\cdots i_{2n}}(\bm{p}_{1},\cdots,\bm{p}_{2n}),
\end{align}
%EEEEEEEEEEEEEEEEEEEEEEEEEEEEEEEEEEEEEEEEEEEE%
 and $C_{i_1\cdots i_{2n}}$ is a $2n$-order cumulant of the
  displacement field in Fourier space $\tilde{\bm{\varPsi}}$, defined
  by
%EEEEEEEEEEEEEEEEEEEEEEEEEEEEEEEEEEEEEEEEEEEE%
\begin{equation}
  \left\langle \tilde{\varPsi}_{i_1}(\bm{p}_1) \cdots
       \tilde{\varPsi}_{i_{2n}}(\bm{p}_{2n}) \right\rangle_{\rm c}
  = (-1)^{n-1}(2\pi)^3 \delta_{\rm D}^3\left(\sum_{i=1}^{2n}\bm{p}_{i}\right)
  C_{i_1\cdots i_{2n}}(\bm{p}_{1},\cdots,\bm{p}_{2n}).
\label{eq:displcumul}
\end{equation}
%EEEEEEEEEEEEEEEEEEEEEEEEEEEEEEEEEEEEEEEEEEEE%

The square of eq.~\eqref{eq:FourierPDF} is naturally factorized out in
LPT for the nonlinear power spectrum. On the other hand, the LPT and
SPT should give the same result when the perturbations are completely
expanded in Eulerian space, since the fundamental equations of motion,
and small perturbations of the expansion are the same in both
perturbation schemes \cite{mat11}. Thus, when we factorize out the
square of eq.~\eqref{eq:FourierPDF} from SPT series of the power
spectrum, we have the following identity:
%EEEEEEEEEEEEEEEEEEEEEEEEEEEEEEEEEEEEEEEEEEEE%
 \begin{multline} \label{LPT_SPT_real}
     P(k)=\exp{\left[-2\sum_{n=1}^{\infty}\frac{k_{i_{1}}\cdots
             k_{i_{2n}}}{(2n)!}A_{i_{1}\cdots i_{2n}}^{(2n)} \right]}
 \\ \times \left\{ P_{\rm L}(k)+ \sum_{m=1}^{\infty}P^{m\mbox{\scriptsize -loop}}_{\rm
       SPT}(k) \right\}
     \left[1
         +\sum_{l=1}^{\infty}\frac{1}{l!}\left(2\sum_{n=1}^{\infty}\frac{k_{i_{1}}\cdots
               k_{i_{2n}}}{(2n)!}A_{i_{1}\cdots i_{2n}}^{(2n)}
         \right)^{l} \right],
\end{multline}
%EEEEEEEEEEEEEEEEEEEEEEEEEEEEEEEEEEEEEEEEEEEE%
where $P^{m\mbox{\scriptsize -loop}}_{\rm SPT}(k)$ is an $m$-loop
contribution to the power spectrum in SPT.
  The indices $i,j,i_1, i_2$ etc.~correspond to spatial components
  and run over $1,2,3$. Repeated appearance of $i_1,i_2,\ldots$
  implicitly means those indices should be summed over. 

  As we mentioned, in the Lagrangian resummation method, the exponential prefactor is kept
  unexpanded, and perturbative truncations are made in the exponent.
  The quantity in the last bracket is normally expanded and truncated
  as usual. In the $N$-loop approximation of  the Lagrangian resummation, the
  series in the exponent is expanded up to ${\cal O}(P_{\rm L})^N$,
  and the remaining factor is expanded up to ${\cal O}(P_{\rm
    L})^{N+1}$. The case of $N=1$ just reduces to
   eq.~\eqref{eq:onelooplpt}.

  In this paper, we generalize eq.~\eqref{eq:onelooplpt} to the case
  of 2-loop approximation, $N=2$. From the consideration above, we do
  not need to calculate LPT from the beginning. Instead, we first
  calculate $P_{\rm SPT}^{m\mbox{\scriptsize -loop}}(k)$ for $m=1,2$,
  and use eq.~\eqref{LPT_SPT_real} truncated at the $N=2$ order 
  to obtain 2-loop
  predictions of the Lagrangian resummation method.

 We need the polyspectra $C_{i_1\cdots i_{2n}}$ of only
  $n=1$, which are given by
%EEEEEEEEEEEEEEEEEEEEEEEEEEEEEEEEEEEEEEEEEEEE%
\begin{align}
C_{ij}(\bm{p},-\bm{p})
  &\equiv C_{ij}(\bm{p}) \notag \\
  &=C^{(11)}_{ij}(\bm{p}) + C^{(22)}_{ij}(\bm{p})
  + C^{(13)}_{ij}(\bm{p})+C^{(31)}_{ij}(\bm{p}),
\end{align}
%EEEEEEEEEEEEEEEEEEEEEEEEEEEEEEEEEEEEEEEEEEEE%
where 
%EEEEEEEEEEEEEEEEEEEEEEEEEEEEEEEEEEEEEEEEEEEE%
\begin{align}
C_{ij}^{(11)}(\bm{p})&=L_{i}^{(1)}(\bm{p})L^{(2)}_{j}(\bm{p})P_{\rm
  L}(p),\\
C_{ij}^{(22)}(\bm{p})&=\frac{1}{2}\int \frac{d^{3}
    p'}{(2\pi)^{3}}L_{i}^{(2)}(\bm{p}',\bm{p}-\bm{p}')L_{j}^{(2)}(\bm{p}',\bm{p}-\bm{p}')P_{\rm
  L}(p')P_{\rm L}(|\bm{p}-\bm{p}'|), \\
C_{ij}^{(13)}(\bm{p})&=C_{ji}^{(31)}(\bm{p}) \notag \\
&=\frac{1}{2}L_{i}^{(1)}(\bm{p})P_{\rm
  L}(p)\int\frac{d^{3}p'}{(2\pi)^{3}}
L_{j}^{(3)}(\bm{p},-\bm{p}',\bm{p}')P_{\rm L}(p'),
\label{eq:cij13}
\end{align}
%EEEEEEEEEEEEEEEEEEEEEEEEEEEEEEEEEEEEEEEEEEEE%
up to ${\cal O}(P_{\rm L})^2$.
The perturbative kernels in LPT up to third order are given by \cite{cat95,CT96}
%EEEEEEEEEEEEEEEEEEEEEEEEEEEEEEEEEEEEEEEEEEEE%
\begin{align}
L_{i}^{(1)}(\bm{p})=&\,\,\frac{k_{i}}{k^{2}}, \label{LPT_kernel1} \\
L^{(2)}_{i}(\bm{p}_{1},\bm{p}_{2}) =&\,\,
\frac{3}{7}\frac{k_{i}}{k^{2}}\left[1-\left(\frac{\bm{p}_{1}\cdot\bm{p}_{2}}{p_{1}p_{2}}\right)^{2}\right], \label{LPT_kernel2} \\ 
L_{i}^{(3{\rm a})}(\bm{p}_{1},\bm{p}_{2},\bm{p}_{3})=&\,\,\frac{5}{7}\frac{k_{i}}{k^{2}}\left[ 1-\left(\frac{\bm{p}_{1}\cdot \bm{p}_{2}}{p_{1}p_{2}}\right)^{2}\right]\left\{ 1-\left[ \frac{(\bm{p}_{1}+\bm{p}_{2})\cdot \bm{p}_{3}}{|\bm{p}_{1}+\bm{p}_{2}|p_{3}}\right]^{2}\right\} \notag \\
&-\frac{1}{3}\frac{k_{i}}{k^{2}}\left[1-3\left(\frac{\bm{p}_{1}\cdot\bm{p}_{2}}{p_{1}p_{2}}\right)^{2} +2\frac{(\bm{p}_{1}\cdot\bm{p}_{2})(\bm{p}_{2}\cdot\bm{p}_{3})(\bm{p}_{3}\cdot\bm{p}_{1})}{p_{1}^{2}p_{2}^{2}p_{3}^{2}} \right] \notag \\
&+\bm{k}\times\bm{T}(\bm{p}_{1},\bm{p}_{2},\bm{p}_{3}), \label{LPT_kernel3a}
\end{align}
%EEEEEEEEEEEEEEEEEEEEEEEEEEEEEEEEEEEEEEEEEEEE%
where $\bm{k}=\bm{p}_{1}+\cdots+\bm{p}_{n}$ for $L_{i}^{(n)}$, and a
vector $\bm{T}$ represents a transverse part whose expression
is not needed in the following application. The third-order
  kernel $\bm{L}^{(3)}$ in eq.~\eqref{eq:cij13} is a symmetrized one
in terms of their arguments:
%EEEEEEEEEEEEEEEEEEEEEEEEEEEEEEEEEEEEEEEEEEEE%
\begin{align}
L_{i}^{(3)}(\bm{p}_{1},\bm{p}_{2},\bm{p}_{3})=\frac{1}{3}\left[
    L^{(3{\rm a})}_{i}(\bm{p}_{1},\bm{p}_{2},\bm{p}_{3}) + {\rm perm.} \right]. \label{LPT_kernel3}
\end{align}
%EEEEEEEEEEEEEEEEEEEEEEEEEEEEEEEEEEEEEEEEEEEE%

Next consider the power spectrum in redshift space. 
In the Lagrangian picture, the displacement field in redshift space,
${\bm{\Psi}}^{\rm s}$,  is distorted by the peculiar velocities, and 
mapped from that in real space as 
%EEEEEEEEEEEEEEEEEEEEEEEEEEEEEEEEEEEEEEEEEEEE%
\begin{align}
{\bm{\Psi}}^{\rm s}=\bm{\Psi}+\frac{\hat{\bm{z}}\cdot\dot{\bm{\Psi}}}{H}\hat{z},
\label{eq:z-r_mapping}
\end{align}
%EEEEEEEEEEEEEEEEEEEEEEEEEEEEEEEEEEEEEEEEEEEE%
where $H$ is is the Hubble parameter, the unit vector $\hat{\bm{z}}$ indicates the line-of-sight direction and $\dot{\bm{\Psi}}=d\bm{\Psi}/dt$.
Using the fact that the time-dependence of the perturbative kernel in real 
space is approximately described by $\Psi^{(n)}\propto D^{n}$ with $D$ being the 
linear growth factor, the displacement field 
of each perturbation order in redshift space is given by
%EEEEEEEEEEEEEEEEEEEEEEEEEEEEEEEEEEEEEEEEEEEE%
\begin{align}
\bm{\Psi}^{{\rm s}(n)}=\bm{\Psi}^{(n)}+nf\left(\hat{\bm{z}}\cdot\bm{\Psi}^{(n)}\right)\hat{\bm{z}},
\label{eq:z-space_mapping}
\end{align}
%EEEEEEEEEEEEEEEEEEEEEEEEEEEEEEEEEEEEEEEEEEEE%
where $f=d\ln{D}/d\ln{a}$ is the logarithmic derivative of the linear 
growth factor. 
Then, the mappings from real space to redshift space of the
order-by-order cumulants are given by
%EEEEEEEEEEEEEEEEEEEEEEEEEEEEEEEEEEEEEEEEEEEE%
\begin{align}
C^{{\rm s}(nm)}_{ij}&=R^{(n)}_{ik}R^{(m)}_{jl}C_{kl}^{(mn)},
\label{eq:mapcumul}
\end{align}
%EEEEEEEEEEEEEEEEEEEEEEEEEEEEEEEEEEEEEEEEEEEE%
where
%EEEEEEEEEEEEEEEEEEEEEEEEEEEEEEEEEEEEEEEEEEEE%
\begin{align}
R^{(n)}_{ij}=\delta_{ij}+nf\hat{z}_{i}\hat{z}_{j}.
\end{align}
%EEEEEEEEEEEEEEEEEEEEEEEEEEEEEEEEEEEEEEEEEEEE%
Thanks to the linear mapping (\ref{eq:z-space_mapping}), 
the perturbation calculation of the displacement field 
is rather straightforward in redshift space, provided the 
perturbative solutions in real-space. 
Nevertheless, the resultant form of the redshift-space power spectrum is 
rather complicated because of the coupling between velocity and 
density fields. The general expressions for redshift-space 
power spectrum is formally written as
%EEEEEEEEEEEEEEEEEEEEEEEEEEEEEEEEEEEEEEEEEEEE%
\begin{multline}\label{LPT_SPT_red}
    P_{\rm s}(\bm{k})=\exp{\left[-2\sum_{n=1}^{\infty}\frac{k_{i_{1}}\cdots
            k_{i_{2n}}}{(2n)!}A_{i_{1}\cdots i_{2n}}^{{\rm s}(2n)}
      \right]}
\\ \times \left\{ (1+f\mu^{2})^{2}P_{\rm L}(k) + \sum_{m=1}^{\infty}P^{m\mbox{\scriptsize
        -loop}}_{\rm sSPT}({\bm{k}}) \right\} 
        \\ \times \left[1
        +\sum_{l=1}^{\infty}\frac{1}{l!}\left(2\sum_{n=1}^{\infty}\frac{k_{i_{1}}\cdots
              k_{i_{2n}}}{(2n)!}A_{i_{1}\cdots i_{2n}}^{{{\rm
                  s}}(2n)} \right)^{l} \right],
\end{multline}
%EEEEEEEEEEEEEEEEEEEEEEEEEEEEEEEEEEEEEEEEEEEE%
where $P^{m\mbox{\scriptsize -loop}}_{\rm sSPT}(k)$ is an
$m$-loop contribution to the redshift-space power spectrum in SPT. 
The quantity $A^{{\rm s}(2n)}_{i_{1}\cdots
  i_{2n}}$ is defined by
%EEEEEEEEEEEEEEEEEEEEEEEEEEEEEEEEEEEEEEEEEEEE%
\begin{align}
    A^{{\rm s}(2n)}_{i_{1}\cdots i_{2n}}= \left[\prod_{i=1}^{2n}\int
        \frac{d^{3}p_{i}}{(2\pi)^{3}}\right] \delta^{3}_{\rm
      D}\left(\sum_{i=1}^{2n}\bm{p}_{i}\right)C_{i_1\cdots
      i_{2n}}^{\rm s}(\bm{p}_{1},\cdots,\bm{p}_{2n}),
\end{align}
%EEEEEEEEEEEEEEEEEEEEEEEEEEEEEEEEEEEEEEEEEEEE%
where $C^{\rm s}_{i_1\cdots i_{2n}}$ is defined by a similar mapping as
that of eq.~\eqref{eq:mapcumul}.

Although the above result involves many additional terms compared to the 
real-space power spectrum, 
several important terms in characterizing redshift distortions 
is identified in eq.~(\ref{LPT_SPT_red}). 
One is the enhancement factor, $(1+f\mu^{2})^{2}$, known as the linear 
Kaiser factor \cite{kai87}. The corrections to the exponential 
prefactor give an additional damping of the power spectrum 
amplitude, which roughly corresponds to the Finger-of-God effect driven 
by the random motion of fluid element. In section~\ref{subsec:redshift_space}, 
we will explicitly see the 
corrections characterizing the redshift distortion up to the 2-loop order.

%==================================================%
\section{Analytic expressions of the 2-loop power spectrum in
  Lagrangian resummation method  \label{sec:analytic}}
%==================================================%
\indent 

Based on the relation
between the power spectrum in the SPT and that in the LPT, 
the 2-loop LPT power spectrum can be expressed as a combination of  1-
and 2-loop SPT power spectra. 
In this section, we derive analytic expressions of the real- and 
redshift-space power spectra up to the 2-loop order in LPT. 

%==================================================%
\subsection{Real space \label{sec:real}}
%==================================================%

Let us first evaluate the power spectrum in real space. The only task 
is to calculate the terms in the exponent
of eq.~\eqref{LPT_SPT_real} up to the demanding order in
$P_{\rm L}(k)$. Up to the second order in $P_{\rm L}(k)$, the exponent of eq.
\eqref{LPT_SPT_real} is expanded as
  \footnote{$A^{(4)}_{ijlm}$ does not have a ${\cal O}(P_{\rm L})^{2}$ connected cumulant.}
%EEEEEEEEEEEEEEEEEEEEEEEEEEEEEEEEEEEEEEEEEEEE%
\begin{align}
-2\sum_{n=1}^{\infty}\frac{k_{i_{1}}\cdots k_{i_{2n}}}{(2n)!}A_{i_{1}\cdots i_{2n}}^{(2n)}& = -k_{i}k_{j} A^{(2)}_{ij} -\frac{1}{12}k_{i}k_{j}k_{l}k_{m}A^{(4)}_{ijlm} + \cdots \notag \\
&\simeq-\frac{k^{2}}{6\pi^{2}} \left[ {\cal A}^{(11)} + {\cal A}^{(22)}+2{\cal A}^{(13)} \right].
\end{align}
%EEEEEEEEEEEEEEEEEEEEEEEEEEEEEEEEEEEEEEEEEEEE%
The quantity ${\cal A}^{(\alpha\beta)}$ is 
calculated from the polyspectra $C^{(\alpha\beta)}_{ ij}$ as
%EEEEEEEEEEEEEEEEEEEEEEEEEEEEEEEEEEEEEEEEEEEE%
\begin{align}
{\cal A}^{(\alpha\beta)}&=
%6\pi^{2}\int \frac{d^{3}{\tmc p}}{(2\pi)^{3}} C^{(\alpha\beta)}_{ii}(\bm{p})
\int_{0}^{\infty} p^{2}dp~{\cal C}^{(\alpha\beta)}(p),
 \end{align} 
%EEEEEEEEEEEEEEEEEEEEEEEEEEEEEEEEEEEEEEEEEEEE%
where we define $C_{ij}^{(\alpha\beta)}(\bm{p})\equiv {\cal
  C}^{\alpha\beta}(p)\hat{p}_{i}\hat{p}_{j}$ and $\hat{p}_i=\bm{p}_i/p$.  
The explicit expressions for ${\cal C}^{\alpha\beta}$ are presented in 
eqs.~(A9)--(A11) of ref.~\cite{mat08a}. In the above, we 
used the fact that $\int
d^{3}p/(2\pi)^{3}\hat{p}_{i}\hat{p}_{j}f(p)=\delta_{ij}/(6\pi^{2})\int
p^{2}dpf(p)$ for an arbitrary function $f(p)$.

Based on the perturbative kernels in LPT defined by eqs.~\eqref{LPT_kernel1}-\eqref{LPT_kernel3}, the 1-loop correction
term, ${\cal A}^{(11)}$, is explicitly written as \cite{mat08a}
%EEEEEEEEEEEEEEEEEEEEEEEEEEEEEEEEEEEEEEEEEEEE%
\begin{align}
{\cal A}^{(11)}&= \int_{0}^{\infty} dp~P_{\rm L}(p).
\end{align}
%EEEEEEEEEEEEEEEEEEEEEEEEEEEEEEEEEEEEEEEEEEEE%
The next-to-leading order contributions, 
${\cal A}^{(22)}$ and ${\cal A}^{(13)}$, are also given by
%EEEEEEEEEEEEEEEEEEEEEEEEEEEEEEEEEEEEEEEEEEEE%
\begin{align}
{\cal A}^{(22)}&=\int_{0}^{\infty} p^{2}dp~{\cal C}^{(22)}(p) \notag \\
&= \int_{0}^{\infty} p^{2}dp \frac{1}{2} \int \frac{d^{3}p_{1}}{(2\pi)^{3}}\delta_{ij} \left[ L_{i}^{(2)}(\bm{p}_{1},\bm{p}-\bm{p}_{1})\right]  \left[ L_{j}^{(2)}(\bm{p}_{1},\bm{p}-\bm{p}_{1})\right] P_{\rm L}(p_{1})P_{\rm L}(|\bm{p}-\bm{p}_{1}|) \notag \\
&=\frac{9}{392\pi^{2}}\int_{0}^{\infty} p_{1}p_{2} dp_{1}dp_{2}P_{\rm L}(p_{1})P_{\rm L}(p_{2}) K\left[ (p_{1}/p_{2}+p_{2}/p_{1})/2 \right], \\
{\cal A}^{(13)}&=\int_{0}^{\infty} p^{2}dp~{\cal C}^{(13)}(p) \notag \\
&= \int_{0}^{\infty} p^{2}dp \frac{1}{2} \delta_{ij} L_{i}^{(1)}(\bm{p})P_{\rm
  L}(p) \int \frac{d^{3}p_{1}}{(2\pi)^{3}} L_{j}^{(3)}(\bm{p},-\bm{p}_{1},\bm{p}_{1})P_{\rm L}(p_{1})  \notag \\
&=\frac{5}{84\pi^{2}}\int_{0}^{\infty} p_{1}p_{2} dp_{1}dp_{2}P_{\rm L}(p_{1})P_{\rm L}(p_{2}) K\left[ (p_{1}/p_{2}+p_{2}/p_{1})/2 \right],
\end{align}
%EEEEEEEEEEEEEEEEEEEEEEEEEEEEEEEEEEEEEEEEEEEE%
where the function $K(y)$ is defined by
%EEEEEEEEEEEEEEEEEEEEEEEEEEEEEEEEEEEEEEEEEEEE%
\begin{align}
K(y)=\frac{5}{3}y-y^{3}-\frac{(y^{2}-1)^{2}}{2}\ln{\left| \frac{y-1}{y+1} \right|}.
\end{align}
%EEEEEEEEEEEEEEEEEEEEEEEEEEEEEEEEEEEEEEEEEEEE%
We note that the function $K(y)$ is always positive in a range $y>0$. 
Also, the 2-loop corrections, ${\cal A}^{(13)}$ and ${\cal A}^{(22)}$, 
are shown to be positive.

Collecting the 2-loop corrections given above, the power spectrum 
up to 2-loop order in LPT is expressed as
%EEEEEEEEEEEEEEEEEEEEEEEEEEEEEEEEEEEEEEEEEEEE%
\begin{align}\label{pk_real}
    P^{2\mbox{\scriptsize -loop}}_{\rm LPT}(k) =&\,\,\exp{\left[ -\frac{k^{2}}{6\pi^{2}}\left({\cal A}^{\rm (11)} + {\cal A}^{\rm (22)}+2{\cal A}^{(13)} \right)\right] }\notag  \\
   & \times \left[ P_{\rm L}(k)+P_{\rm SPT}^{1\mbox{\scriptsize
            -loop}}(k) + P_{\rm SPT}^{2\mbox{\scriptsize -loop}}(k)  \right.  \notag \\
   & \left.~~+ P_{\rm L}(k)\left\{\frac{k^{2}}{6\pi^{2}}\left({\cal
                  A}^{\rm (11)} + {\cal A}^{\rm (22)} +2{\cal
                  A}^{(13)} \right) + \frac{k^{4}}{72\pi^{4}}\left({\cal A}^{(11)}\right)^{2}
        \right\} + \frac{k^{2}}{6\pi^{2}}P_{\rm
          SPT}^{1\mbox{\scriptsize -loop}}(k){\cal A}^{\rm(11)}
    \right],
\end{align}
%EEEEEEEEEEEEEEEEEEEEEEEEEEEEEEEEEEEEEEEEEEEE%
%and $9/392+2\times5/84=167/1176$.
where $P^{1\mbox{\scriptsize -loop}}_{\rm SPT}(k)$ and
$P^{2\mbox{\scriptsize -loop}}_{\rm SPT}(k)$ are 1- and 2-loop
contributions to the real space power spectrum in SPT respectively.

%sssssssssssssssssssssssssssssssssssssssssssssssssssssssssss%
\subsection{Redshift space \label{sec:red}}
\label{subsec:redshift_space}
%sssssssssssssssssssssssssssssssssssssssssssssssssssssssssss%

Next consider the power spectrum in observable redshift space. In
similar manner to the real space, we start with 
eq.~\eqref{LPT_SPT_red} by expanding the exponent 
up to the second order in $P_{\rm L}(k)$: 
%EEEEEEEEEEEEEEEEEEEEEEEEEEEEEEEEEEEEEEEEEEEE%
\begin{align}\label{expand_red}
-2\sum_{n=1}^{\infty}\frac{k_{i_{1}}\cdots k_{i_{2n}}}{(2n)!}A_{i_{1}\cdots i_{2n}}^{{\rm s}(2n)}& = -k_{i}k_{j} A^{{\rm s}(2)}_{ij} -\frac{1}{12}k_{i}k_{j}k_{l}k_{m}A^{{\rm s}(4)}_{ijlm} + \cdots \notag \\
&\simeq- \frac{k_ik_j}{6\pi^{2}} \left[ {\cal A}_{ij}^{{\rm s}(11)} +{\cal  A}_{ij}^{{\rm s}(22)} +{\cal A}_{ij}^{{\rm s}(13)}+{\cal A}_{ij}^{{\rm s}(31)} \right].
\end{align}
%EEEEEEEEEEEEEEEEEEEEEEEEEEEEEEEEEEEEEEEEEEEE%
The quantity ${\cal A}^{{\rm s}(\alpha\beta)}$ 
represents a momentum integration of a polyspectrum in redshift space.
Based on the relation between the polyspectra in real and redshift spaces, 
we obtain 
%EEEEEEEEEEEEEEEEEEEEEEEEEEEEEEEEEEEEEEEEEEEE%
\begin{align}
{\cal A}^{{\rm s}(\alpha\beta)}_{ij}&= 6\pi^{2}\int \frac{d^{3} p}{(2\pi)^{3}} C^{{\rm s}(\alpha\beta)}_{ij}(\bm{p}) \notag \\
&=\left\{\delta_{ij}+(\alpha +\beta)f\hat{z}_{i}\hat{z}_{j}+\alpha\beta f^{2}\hat{z}_{i}\hat{z}_{j}\right\} {\cal A}^{(\alpha\beta)}.
\end{align}
%EEEEEEEEEEEEEEEEEEEEEEEEEEEEEEEEEEEEEEEEEEEE%
Taking an inner product of ${\cal A}^{{\rm
    s}(\alpha\beta)}_{ij}$ with $k_{i}k_{j}$ , we have
%EEEEEEEEEEEEEEEEEEEEEEEEEEEEEEEEEEEEEEEEEEEE%
\begin{align}
k_{i}k_{j}{\cal A}^{{\rm s}(\alpha\beta)}_{ij}=k^{2}{\cal D}^{(\alpha\beta)}(f,\mu){\cal A}^{(\alpha\beta)}.
\end{align}
%EEEEEEEEEEEEEEEEEEEEEEEEEEEEEEEEEEEEEEEEEEEE%
The quantity $\mu=\hat{\bm{z}}\cdot\bm{k}/k$ is the directional 
cosine of the wave vector $\bm{k}$ with respect to the line of sight.
The function ${\cal D}^{(\alpha\beta)}(\mu,f)$ which describes the 
redshift distortion is defined by
%EEEEEEEEEEEEEEEEEEEEEEEEEEEEEEEEEEEEEEEEEEEE%
\begin{align}
{\cal D}^{(\alpha\beta)}(\mu, f)={\cal D}^{(\beta\alpha)}(\mu,f) \equiv1+(\alpha + \beta +  \alpha\beta f)f\mu^{2}.
\end{align} 
%EEEEEEEEEEEEEEEEEEEEEEEEEEEEEEEEEEEEEEEEEEEE%
From these relations, we can get the non-linear power spectrum in
redshift space up to 2-loop order.
The result is
%EEEEEEEEEEEEEEEEEEEEEEEEEEEEEEEEEEEEEEEEEEEE%
\begin{align}\label{pk_red}
P^{2\mbox{\scriptsize -loop}}_{\rm sLPT}(\bm{k}) =&\,\,\exp{\left[ -\frac{k^{2}}{6\pi^{2}}\left\{{\cal D}^{(11)}{\cal A}^{\rm (11)} + {\cal D}^{(22)}{\cal A}^{\rm (22)}  + 2 {\cal D}^{(13)}{\cal A}^{\rm (13)} \right\}\right] } \notag \\
 &\times \left[ (1+f\mu^{2})^{2}P_{\rm L}(k)+P_{\rm sSPT}^{1\mbox{\scriptsize -loop}}(\bm{k}) + P_{\rm sSPT}^{2\mbox{\scriptsize -loop}}(\bm{k})  \right. \notag \\ 
 &\left.~~+ (1+f\mu^{2})^{2}P_{\rm L}(k) \left\{ \frac{k^{2}}{6\pi^{2}}\left({\cal D}^{(11)}{\cal A}^{\rm (11)} + {\cal D}^{(22)}{\cal A}^{\rm (22)} +2{\cal D}^{(13)}{\cal A}^{(13)} \right) \right.\right. \notag \\
& \left.\left.~~+ \frac{k^4}{72\pi^{2}}\left({\cal D}^{(11)}{\cal A}^{(11)}\right)^{2}  \right\}+  \frac{k^{2}}{6\pi^{2}}P_{\rm sSPT}^{1\mbox{\scriptsize -loop}}(\bm{k}){\cal D}^{(11)}{\cal A}^{\rm(11)} \right].
\end{align}
%EEEEEEEEEEEEEEEEEEEEEEEEEEEEEEEEEEEEEEEEEEEE%
where $P^{1\mbox{\scriptsize -loop}}_{\rm sSPT}(\bm{k})$ and
$P^{2\mbox{\scriptsize -loop}}_{\rm sSPT}(\bm{k})$ are 1-
and 2-loop contributions to the redshift-space 
power spectrum in SPT, respectively.

  In eq.~(\ref{pk_red}), the terms proportional to $f\,\mu^2$ 
  represent the corrections characterizing the redshift distortions. 
  These corrections appear not only in the spatial-correlation terms, which 
  have been expanded from the exponent, but also in the exponential 
  prefactor. Thus, the resultant redshift-space power spectrum 
  naturally possesses an additional damping effect 
  responsible for the Finger-of-God (FoG) effect. Note that 
  a naive perturbative treatment based on the SPT usually 
  fails to describe the FoG effect, 
  which should be phenomenologically introduced by hand (but 
  see \cite{taruya10}).  By contrast, 
  in LPT, the redshift distortion is described by a 
  simple linear mapping from real space, (\ref{eq:z-r_mapping}) or 
  (\ref{eq:z-space_mapping}), and 
  with the resummation method, the power spectrum damping by the 
  FoG effect is naturally incorporated into the redshift-space power 
  spectrum in a non-perturbative manner.

While the resultant expression for redshift-space power spectrum 
seems to contain physically relevant effects of non-linear redshift 
distortions, an explicit evaluation of eq.~(\ref{pk_red}) needs 
a bit technical aspect. To be specific, 
a difficult part is the term $P^{2\mbox{\scriptsize -loop}}_{\rm sSPT}(\bm{k})$.  
It involves six-dimensional integral, and 
is characterized as a function of $k$ and directional cosine $\mu$. 
No one has so far tried to evaluate it, and at present, 
the explicit calculation of $P^{2\mbox{\scriptsize -loop}}_{\rm sLPT}(\bm{k})$ 
seems infeasible unless we develop an efficient technique to compute 
$P^{2\mbox{\scriptsize -loop}}_{\rm sSPT}(\bm{k})$.  Hence, 
we leave the discussion on the numerical evaluation of the redshift-space 
power spectrum for future work, and 
we hereafter focus on the quantitative prediction of 
real-space power spectrum. 

\subsection{Evaluations of the 1- and 2-loop power spectra in SPT}

The expression of real-space power spectrum given 
in eq.~(\ref{pk_real})  involves
the 1- and 2-loop power spectra in SPT, and we 
need to directly evaluate them for a quantitative prediction of the power 
spectrum. Here, we briefly explain how to compute these two contributions. 
The explicit expressions for the 1- and 2-loop corrections in SPT are 
decomposed into several pieces as 
%EEEEEEEEEEEEEEEEEEEEEEEEEEEEEEEEEEEEEEEEEEEE%
\begin{align}
&P_{\rm SPT}^{1\mbox{\scriptsize-loop}}(k)=P_{\rm SPT}^{(22)}(k)+P_{\rm SPT}^{(13)}(k),
\nonumber\\
&P_{\rm SPT}^{2\mbox{\scriptsize-loop}}(k)=P_{\rm SPT}^{(33)}(k)+
P_{\rm SPT}^{(24)}(k)+P_{\rm SPT}^{(15)}(k),
\end{align}
%EEEEEEEEEEEEEEEEEEEEEEEEEEEEEEEEEEEEEEEEEEEE%
where the quantities $P_{\rm SPT}^{(mn)}$ imply the ensemble averages 
obtained from the $m$-th and $n$-th order perturbative solutions of SPT. 
They are expressed as a collection of multi-dimensional integrals involving 
the perturbative kernel of the higher-order solutions 
(e.g., \cite{Fry94,Carlson09,Taruya09}): 
%%%%%%%%%%%%%%%%%%%%%%%%%%%%%%%%%%%%%%%%%%%%%%%%%%%%%%%
\begin{align}
P_{\rm SPT}^{(22)}(k) =& \,\,2\,\,\int \frac{d^3q}{(2\pi)^3}\,
\left\{ {\cal F}^{(2)}(\bm{q},\bm{k}-\bm{q}) \right\}^{2} \,
P_{\rm L}(q)P_{\rm L}(|\bm{k}-\bm{q}|),
\\
P_{\rm SPT}^{(13)}(k)=& \,\,6 \,P_{\rm L}(k)\,\int \frac{d^3q}{(2\pi)^3}\,
{\cal F}^{(3)}(\bm{k},\bm{q},-\bm{q}) \,P_{\rm L}(q),
\\
P_{\rm SPT}^{(33)}(k)=& \,\,9 \,P_{\rm L}(k)\,\left\{ \int \frac{d^3p}{(2\pi)^3}\,
{\cal F}^{(3)}(\bm{k},\bm{p},-\bm{p})  \,P_{\rm L}(p) \right\}^{2}
\nonumber\\
&+ 6 \,\int \frac{d^3p d^3q}{(2\pi)^6}\,
\left\{{\cal F}^{(3)}(\bm{p},\bm{q},\bm{k}-\bm{p}-\bm{q}) \right\}^2
\,P_{\rm L}(p)\,P_{\rm L}(q)\,P_{\rm L}(|\bm{k}-\bm{p}-\bm{q}|), 
\\
P_{\rm SPT}^{(24)}(k)=& \,\,24\int \frac{d^3p d^3q}{(2\pi)^6}\,
{\cal F}^{(2)}(\bm{p}, \bm{k}-\bm{p}) \,
{\cal F}^{(4)}(\bm{p},\bm{q},-\bm{q},\bm{k}-\bm{p})\,
P_{\rm L}(p)\,P_{\rm L}(q)\,P_{\rm L}(|\bm{k}-\bm{p}|),
\end{align}
\begin{align}
P_{\rm SPT}^{(15)}(k)=& \,\,30 P_{\rm L}(k)\,\int \frac{d^3p d^3q}{(2\pi)^6}\,
{\cal F}^{(5)}(\bm{p},\bm{q},\bm{k},-\bm{p},-\bm{q})\,\,
P_{\rm L}(p)\,P_{\rm L}(q).
\end{align}
%%%%%%%%%%%%%%%%%%%%%%%%%%%%%%%%%%%%%%%%%%%%%%%%%%%%%%%
Here, ${\cal F}^{(n)}(\bm{k}_1,\cdots,\bm{k}_n)$ is the symmetrized 
kernel of the $n$-th order solutions for the density field 
(see \cite{BCGS02, Taruya09} for derivation and explicit expressions).

Note that the expressions of the 1-loop power spectra 
$P_{\rm SPT}^{1\mbox{\scriptsize-loop}}$ 
can be further reduced to   
the one-dimensional and two-dimensional integral for $P^{(13)}$ and 
$P^{(22)}$, respectively (e.g., \cite{SM91, MSS91, JB94}).  
For the explicit computation of power spectrum in next section, 
we use the method of Gaussian quadratures for 
numerical integration of 1-loop power spectra. On the other hand, 
for the 2-loop power spectra, except for the first term in $P^{(33)}$,  
the integrals are no longer simplified, and we directly evaluate 
the five-dimensional integration (due to the symmetry around the vector 
$\bm{k}$, one of the azimuthal angles is trivially integrated). 
In this paper, following the approach by \cite{Taruya09}, 
we adopt the Monte-Carlo technique of quasi-random sampling using the 
library \verb|Cuba|\footnote{\tt http://www.feynarts.de/cuba/}.  
The symmetrized kernels for perturbative solutions higher than the 
third order are too complicated to express analytically, and 
we numerically generated the symmetrized kernels using the recursion 
relation of perturbative solutions (e.g., \cite{Taruya09,GGRW86,CS06a,BCGS02}).

%SSSSSSSSSSSSSSSSSSSSSSSSSSSSSSSSSSSSSSSSSSSSS%
\section{Comparison with $N$-body simulations in real space
\label{sec:compare}}
%SSSSSSSSSSSSSSSSSSSSSSSSSSSSSSSSSSSSSSSSSSSSS%

In this section, in order to quantify the effect of 
the next-to-leading order corrections in LPT, 
we explicitly compute the 2-loop corrections, and 
compare the analytic predictions with $N$-body simulations. 
We adopt a cosmological model with the WMAP 5yr \cite{WMAP5} 
parameters $\Omega_{\rm m}=0.279$, $\Omega_{\rm \Lambda}=0.721$, $\Omega_{\rm b}=0.046$, $h=0.701$, $n_{s}=0.96$ and $\sigma_{8}=0.817$, 
and the linear power spectrum $P_{\rm L}(k)$ is calculated from the 
output of the CAMB code \cite{CAMB}.

%ssssssssssssssssssssssssssssssssssssssssssssssssssssssssssss%
\subsection{$N$-body simulations }
%ssssssssssssssssssssssssssssssssssssssssssssssssssssssssssss%

\begin{table}
\begin{center}
\caption{Parameters of $N$-body simulations.}
\label{Nbody}
\begin{tabular}{lcccc}  \toprule
Name & $L_{\rm box} $ &  \# of particles & $z_{\rm init}$ & \# of runs \\ \midrule
Main & $1,000h^{-1}{\rm Mpc}$ & $512^{3}$ & $31$ & $30$ \\
High (L11-N11) & $2,048 h^{-1}{\rm Mpc}$ & $2048^{3}$ & $99$ & $1$ \\ 
~~~~~~~(L12-N11) & $4,096 h^{-1}{\rm Mpc}$ & $2048^{3}$ & $99$ & $1$ \\ 
\bottomrule
\end{tabular}
\end{center}
\end{table}

To compare the LPT results with $N$-body simulations, we use the 
subset of the $N$-body data presented in \cite{Taruya09}.  
The data were created by a public $N$-body code \verb|GADGET2| 
\cite{Springel05}
with cubic boxes of side length $1\,h^{-1}$Gpc, $512^3$ particles.  
The initial conditions were generated by the \verb|2LPT| code \cite{Crocce06} 
at $z_{\rm init}=31$, and the output data 
of the $30$ independent realizations 
were stored at redshifts $z=3$, $2$, 
$1$, and $0.5$. 
In addition to the main simulations, we also use the results of 
high-resolution simulation data presented in ref.~\cite{VN11}, 
which were similarly created by \verb|GADGET2| with the initial condition 
generator, \verb|2LPT|.
We summarize our sets of $N$-body simulation in Table~\ref{Nbody}.

The power spectra of these simulations were computed by
combining several output results with different box sizes, and
for the purpose to probe BAOs,
we specifically use the results of $N$-body runs called L11-N11 and L12-N11 in
ref.~\cite{VN11}, whose box sizes are  $2,048$ and $4,096\,h^{-1}$
Mpc, respectively.
The simulations were evolved from the redshift $z_{\rm init}=99$ and the
output results were stored at $z=3$ and $1$.  Although the realization
of these simulations is
only one, each simulation contains $2,048^{3}$ particles, and
we preferentially use the L11-N11 run to measure the small-scale power spectrum.

We measure the matter power spectrum adopting 
the same treatment as described in ref.~\cite{Nishimichi09}. 
That is, we calculate the power spectrum from the Fourier transform of 
the density field assigned on the $1,024^3$ grids with the Cloud-in-Cells 
interpolation. To reduce the effect of finite-mode sampling advocated by 
ref.~\cite{Takahashi08}, 
the resultant power spectrum is multiplied by the ratio
$P_{\rm lin}(k)/\widehat{P}^{\rm PT}(k)$ at $k\lesssim0.1h$Mpc$^{-1}$, 
where the quantity 
$\widehat{P}^{\rm PT}(k)$ is calculated from the SPT up to the 
third-order in density field, and $P_{\rm lin}(k)$ is the input linear 
power spectrum extrapolated to a given output redshift. 
In computing $\widehat{P}^{\rm PT}(k)$, we use the Gaussian-sampled 
density field used to generate the initial condition of each $N$-body run 
(see refs.~\cite{Takahashi08,Nishimichi09} in detail). 
As for the estimation of two-point correlation function, we use the 
grid-based calculation using the Fast Fourier Transformation (FFT) 
\cite{Taruya09}. Similar to the 
power spectrum analysis, we first compute the square of the density 
field on each grid of Fourier space. Then, applying the inverse Fourier 
transformation, we take the average over realization, 
and obtain the two-point correlation function. Note that
the finite-mode sampling also affects the calculation of 
the two-point correlation function. We correct it by 
subtracting and adding the extrapolated linear density field as, 
$\widehat{\xi}(r)-\widehat{\xi}_{\rm lin}(r)+\xi_{\rm lin}(r)$,   
where $\widehat{\xi}_{\rm lin}$ is the correlation function estimated 
from the Gaussian density field, and $\xi_{\rm lin}$ is the 
linear theory prediction of two-point correlation function.

%ssssssssssssssssssssssssssssssssssssssssssssssssssssssssssss%
\subsection{Power spectrum}
%ssssssssssssssssssssssssssssssssssssssssssssssssssssssssssss%

%FFFFFFFFFFFFFFFFFFFFFFFFFFFFFFFFFFFFFFFFFFFFFFFFFFF%
\begin{figure}
\centering
\includegraphics[width=110mm,angle=-90]{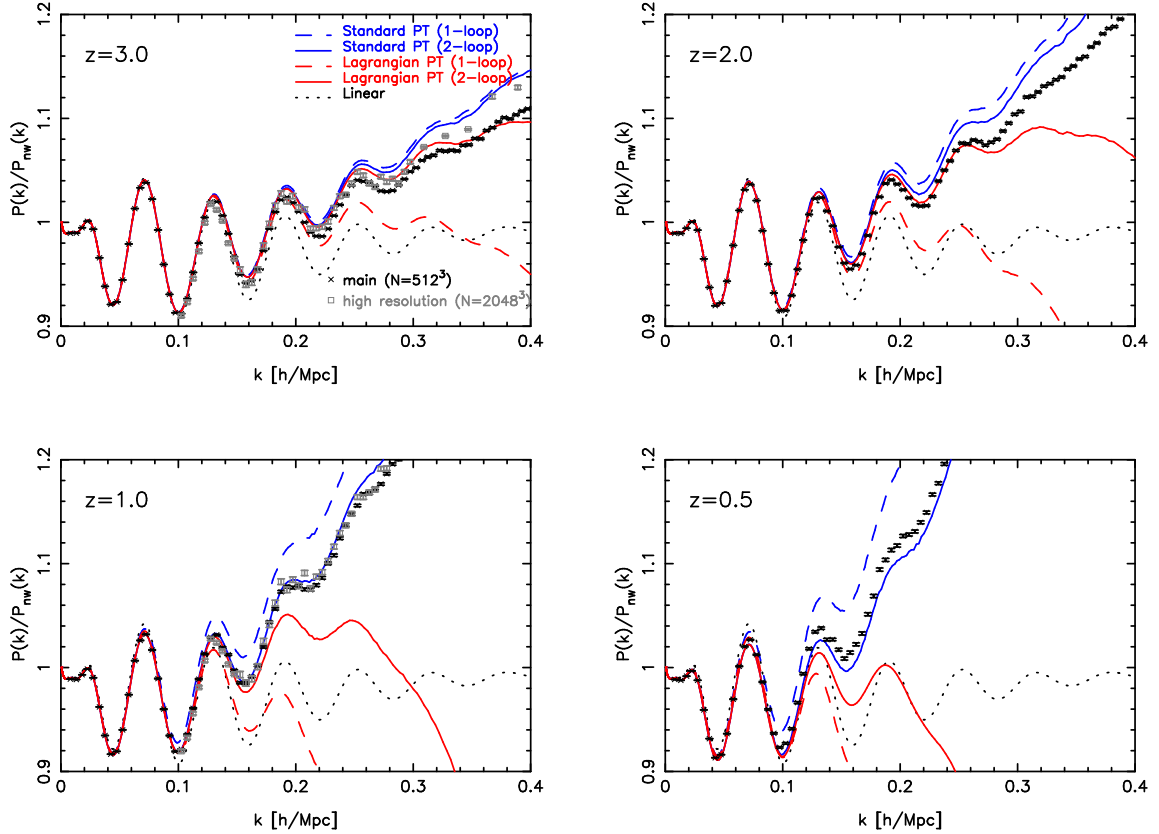}
\caption{Ratio of power spectrum to smoothed reference spectrum, 
$P(k)/P_{\rm nw}(k)$, given at redshifts $z=3$ (top left), 
$2$ (top right), $1$ (bottom left) and $0.5$ (bottom right).
The power spectra from linear theory (black dotted), 1-loop 
SPT (blue dashed), 2-loop SPT (blue solid), 1-loop LPT (red dashed) 
and 2-loop LPT (red solid) are compared with $N$-body simulations.
 Results of the main and high-resolution simulations are shown 
as black and gray symbols with error bars, respectively.
The reference power spectrum $P_{\rm nw}(k)$ is calculated 
from the no-wiggle formula of the linear transfer function 
in ref.~\cite{EH}.}
\label{pk_ratio}
\end{figure}
%FFFFFFFFFFFFFFFFFFFFFFFFFFFFFFFFFFFFFFFFFFFFFFFFFFF%

%FFFFFFFFFFFFFFFFFFFFFFFFFFFFFFFFFFFFFFFFFFFFFFFFFFF%
\begin{figure}
\centering
\includegraphics[width=110mm,angle=-90]{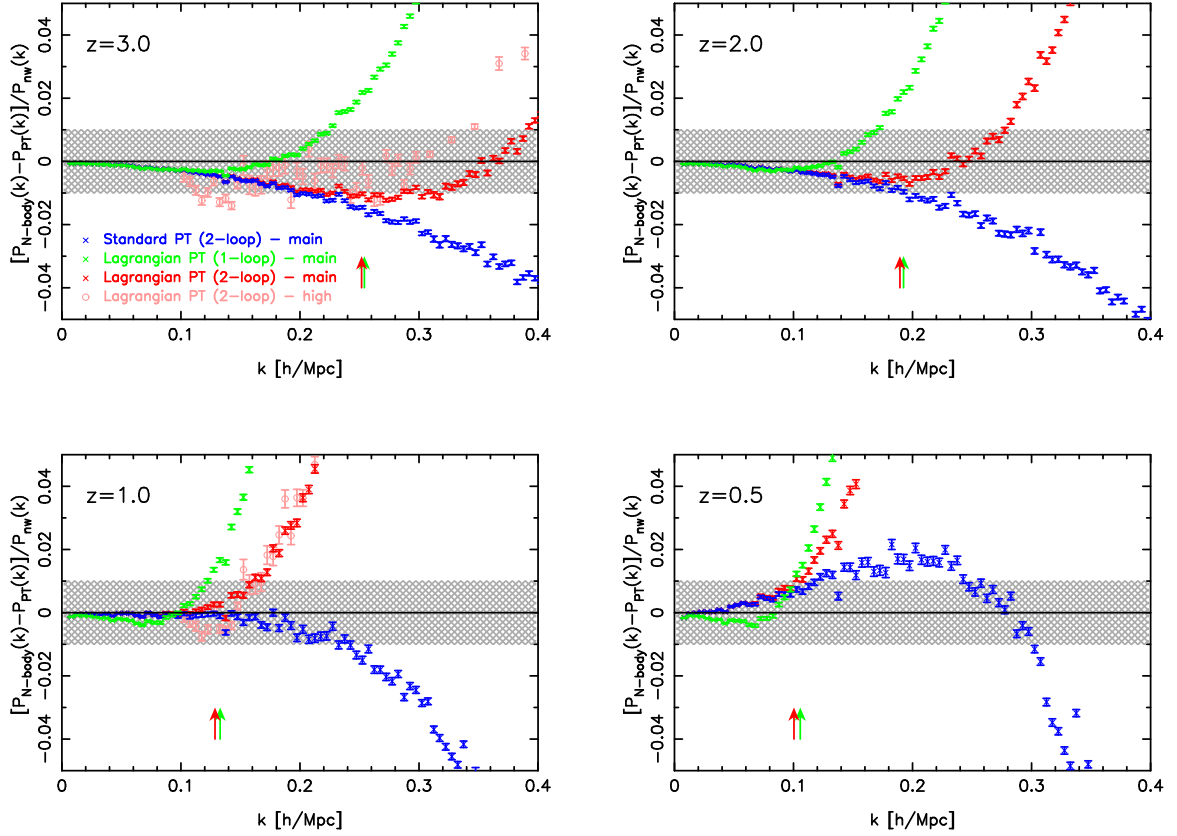}
\caption{Difference between the results of main $N$-body simulation and PT 
(2-loop SPT: blue, 1-loop LPT: green, 2-loop LPT: red) divided by the 
reference spectrum, $[P_{N-{\rm body}}(k)-P_{\rm PT}(k)]/P_{\rm nw}(k)$.
We also show the difference between the result of high-resolution 
simulation and 2-loop LPT with light red symbols.
In each panel, green and red vertical arrows represent the wave 
number $k^{1\mbox{\scriptsize -loop}}_{\rm nl}/2$ and $k^{2\mbox{\scriptsize -loop}}_{\rm nl}/2$, which are respectively estimated from the exponential pre-factor 
in the 1- and 2-loop expressions for real-space power spectrum in LPT 
[see eqs.(\ref{eq:k_nl_1loop}) and (\ref{eq:k_nl_2loop})].}
\label{pk_diff}
\end{figure}
%FFFFFFFFFFFFFFFFFFFFFFFFFFFFFFFFFFFFFFFFFFFFFFFFFFF%

Figure \ref{pk_ratio} shows the ratio of power spectrum 
to a smooth reference spectrum, 
$P(k)/P_{\rm nw}(k)$, 
where the function $P_{\rm nw}(k)$ is the linear power spectrum calculated 
from the smooth transfer function neglecting the BAO feature in 
ref.~\cite{EH}. From top to bottom,    
the results at redshifts $z=3$, $2$, $1$ and $0.5$ are shown. 
In each panel, 
the power spectra from linear theory (black dotted), 1-loop SPT 
(blue dashed), 2-loop SPT (blue solid), 1-loop LPT (red dashed) and 
2-loop LPT (red solid) are plotted, and are compared with $N$-body 
simulations.
Results of the main and high-resolution simulations are 
  respectively shown as black and gray symbols with error bars. 
Note that owing to $30$ independent realizations 
and the correction of the finite-mode sampling 
by ref.~\cite{Takahashi08}, the scatter of the $N$-body results 
is rather reduced, and the size of each error bar becomes 
hard to see visually. 
Overall, the predictions of the 2-loop LPT show better agreement with $N$-body 
simulations. The range of the agreement is wider than that 
of the 1-loop LPT at all redshifts, indicating that 
the 2-loop corrections in LPT can give an important contribution to the 
nonlinear enhancement of the power spectrum at high $k$. 
On the other hand, the 2-loop correction 
in SPT slightly reduces the amplitude of 
power spectrum, and the predictions of 2-loop SPT tend to reproduce 
the $N$-body results quite well compared to the 1-loop SPT. 
However, a closer look at the behaviors on small scales 
reveals that the predictions overestimate the $N$-body results 
at high redshift, and then turn to slightly underestimate at low redshift. 
Note that the power spectra obtained 
from the main and high-resolution simulations show somewhat different 
behaviors at $z=3$. At $k\gtrsim0.2\,h$Mpc$^{-1}$, the result of 
main simulations gets smaller power, and the discrepancy between 
simulation and the 2-loop predictions apparently manifests. Presumably, 
this would be explained by a transient phenomenon due to 
the lack of number of particles. 
Low resolution simulations with a small number of particles 
usually suffer from the discreteness effect,  
and the $N$-body data on the scales below the particle mean separation 
are not reliable particularly at an early time. 
Later, as the gravitational clustering develops, 
the non-linear mode transfer is expected to eventually dominate 
the initial power, and the small-scale structure tend to catch up the 
actual nonlinear growth (e.g., \cite{hamana02}). Although this 
qualitative picture is specifically relevant for the 
small-scales clustering, the apparent lack of initial power 
in low-resolution simulations may be influential even on 
relatively large scales, $k\gtrsim0.2\,h$Mpc$^{-1}$, as 
shown in detail by ref.~\cite{VN11} (see figure~3 of their paper).  
It can affect the estimation of the validity 
range of PT results, and we need to take care of the systematics.

In figure~\ref{pk_diff}, to investigate the agreement 
in more quantitative ways, we show the differences of the power 
spectrum between simulation and PT prediction normalized by the smooth 
reference spectrum, i.e.,
$[P_{N\mbox{\scriptsize -body}}-P_{\rm PT}]/P_{\rm nw}(k)$.
Here, the differences between the main simulation and 
predictions are mainly shown, but we also plot the 
case with high-resolution simulation for 2-loop LPT (
main vs 2-loop SPT: blue, main vs 1-loop LPT: green, 
main vs 2-loop LPT: red, high-resolution vs 2-loop LPT: light red). 
In each panel of figure \ref{pk_diff}, 
the green and red vertical arrows respectively represent 
the characteristic wave numbers $k^{1\mbox{\scriptsize -loop}}_{\rm nl}/2$ and $k^{2\mbox{\scriptsize -loop}}_{\rm nl}/2$ defined by 
%EEEEEEEEEEEEEEEEEEEEEEEEEEEEEEEEEEEEEEEEEEEE%
\begin{align}
k^{1\mbox{\scriptsize -loop}}_{\rm nl}=&\left( \frac{ {\cal A}^{(11)} }{6\pi^{2}}\right)^{-1/2} , \label{eq:k_nl_1loop}
\\
k^{2\mbox{\scriptsize -loop}}_{\rm nl}=&\left[ \frac{ {\cal A}^{(11)}+{\cal A}^{(22)}+
2{\cal A}^{(13)} }{6\pi^{2}}\right]^{-1/2}.
\label{eq:k_nl_2loop}
\end{align}
%EEEEEEEEEEEEEEEEEEEEEEEEEEEEEEEEEEEEEEEEEEEE%
These wave numbers indicate the characteristic damping scales of the 
power spectrum in LPT, which are natural extension of the 
definition of the nonlinear scale, $k_{\rm nl}$, in ref.~\cite{mat08a} 
[see eq.~(38) of this paper].

Compared to the predictions of 1-loop LPT, 
the 2-loop corrections in LPT extend the range of agreement with 
1\% precision by a factor of 1.0 ($z=0.5$), 1.3 ($1$), 1.6 ($2$) 
and 1.8 ($3$). The ongoing (e.g., HETDEX \cite{HETDEX})
and proposed surveys (e.g., BigBOSS \cite{BigBOSS}, WFIRST 
\cite{WFIRST}, and SuMIRe
\cite{SuMIRe}) plan to precisely measure the BAOs at $z=1$--$2$, 
and thus the improvement with the 2-loop correction at
high-$z$ is practically important and encouraging for an accurate 
determination of acoustic scales. 
Figure~\ref{pk_diff} also implies that LPT has a better 
convergence property for the higher-order corrections, since the range of 
agreement becomes monotonically wider as we include the higher-order 
corrections. Apparently, at lower redshifts, the range of agreement of the 
2-loop SPT becomes comparable to that of the 2-loop 
LPT. As seen in figure~\ref{pk_ratio}, however, the 2-loop corrections  
in SPT can give a negative contribution at large $k$, and it is not 
guaranteed that the inclusion of higher-order corrections in SPT 
monotonically improves the prediction. In this respect, the agreement 
in SPT shown at lower redshifts may be regarded as accidental one. 
As the non-linearity significantly develops, even the 2-loop LRT is insufficient to improve the validity range of prediction. 
Although we did not store the simulation data at $z=0$, it is expected that the validity range of 2-loop LPT at $z=0$ becomes rather comparable to that of 1-loop LPT, and the 1\% precision would be achieved only at $k\lesssim0.7 h^{-1}{\rm Mpc}$ (see ref.~\cite{Carlson09}).

Finally, it is worth mentioning a relation between the validity range of 
LPT and the exponential damping scale. 
Ref.~\cite{mat08a} pointed out that the damping scale
is considered as a criteria for the validity range of the
1-loop LPT, and we can see that the $k^{{1\mbox{\scriptsize -loop}}}_{\rm nl}/2$ becomes a plausible indicator of the agreement with $N$-body simulations. On the other hand, 
figure \ref{pk_diff} shows that such a
 criteria does not simply work for the 2-loop LPT. 
Though the damping scale of the 2-loop LPT is always 
smaller than that of the 1-loop LPT, due to the
higher-order corrections in the exponential prefactor, the validity
range of the 2-loop LPT becomes wider than the scale indicated 
by $k^{2\mbox{\scriptsize -loop}}_{\rm nl}/2$ or $k^{{1\mbox{\scriptsize -loop}}}_{\rm nl}/2$. In this sense, 
the damping scale $k^{2\mbox{\scriptsize -loop}}_{\rm nl}/2$ 
may be regarded as the scale where the prediction of 2-loop LPT starts to
deviate from that of the 2-loop SPT.

%===================================================%
\subsection{Correlation function}
%===================================================%
%FFFFFFFFFFFFFFFFFFFFFFFFFFFFFFFFFFFFFFFFFFFFFFFFFFF%
\begin{figure}
\centering
\includegraphics[width=110mm,angle=-90]{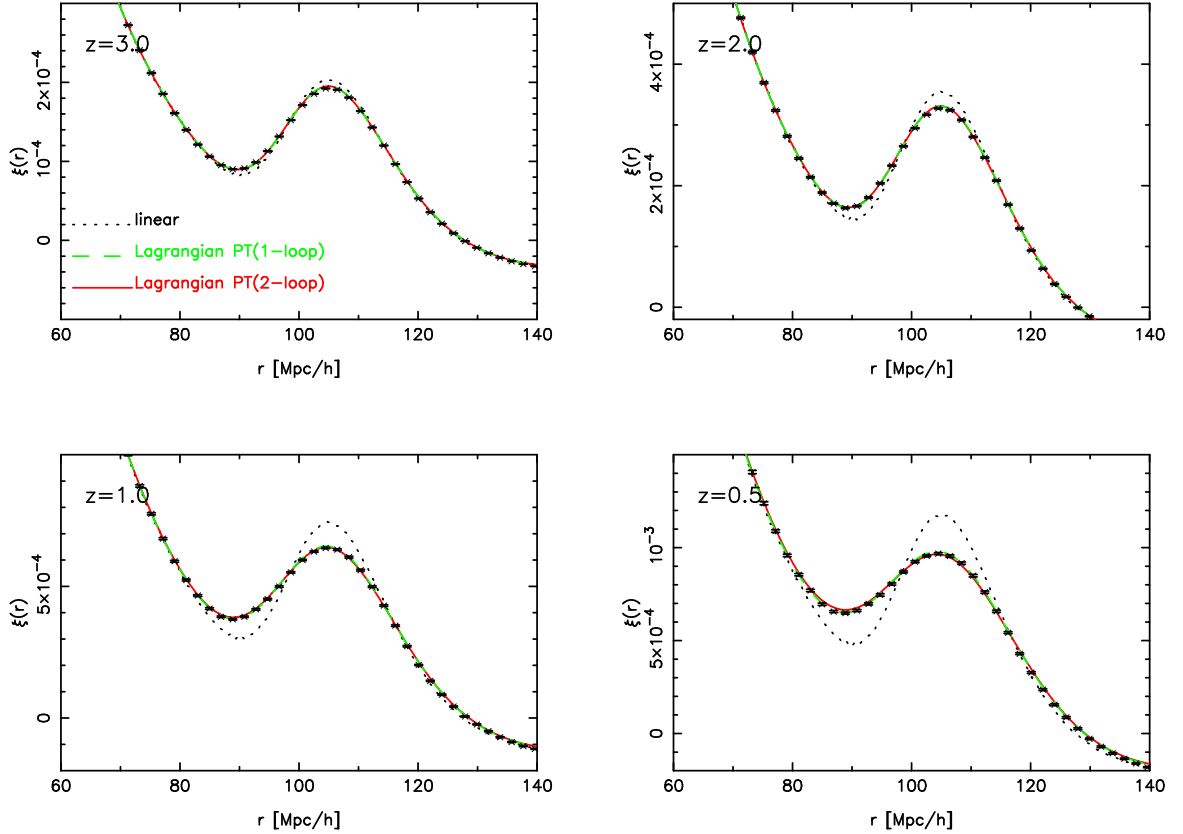}
\caption{Two-point correlation functions given at redshifts $z=3$ (top left), $2$ (top right), $1$ (bottom left) and $0.5$ (bottom right). 
Two-point correlation functions from linear (black dotted), 1-loop LPT (green dashed) and 2-loop LPT (red solid) are compared with $N$-body simulations (black symbol with error bars).}
\label{xi}
\end{figure}
%FFFFFFFFFFFFFFFFFFFFFFFFFFFFFFFFFFFFFFFFFFFFFFFFFFF%

%FFFFFFFFFFFFFFFFFFFFFFFFFFFFFFFFFFFFFFFFFFFFFFFFFFF%
\begin{figure}[t]
\centering
\includegraphics[width=110mm,angle=-90]{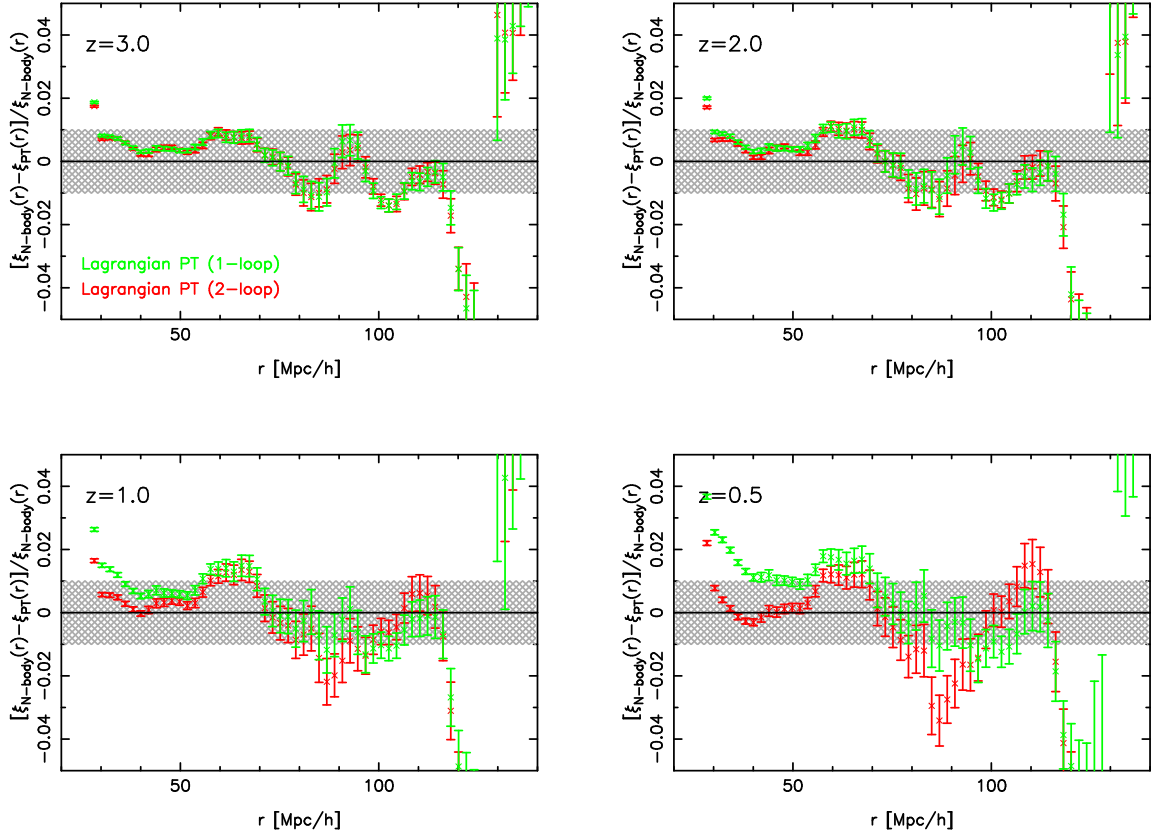}
\caption{Fractional differences between $N$-body and Lagrangian PT results, 
  $[\xi_{N-{\rm body}}(r)-\xi_{\rm PT}(r)]/\xi_{N-{\rm body}}(r)$; 
  1-loop: green, 2-loop: red.}
\label{xi_diff}
\end{figure}
%FFFFFFFFFFFFFFFFFFFFFFFFFFFFFFFFFFFFFFFFFFFFFFFFFFF%

The two-point correlation function $\xi(r)$ is given by 
Fourier transform of the power spectrum:
\begin{align}\label{tpcf}
\xi(r)=\int^{\infty}_{0} \frac{k^{2}dk}{2\pi^{2}}\frac{\sin{(kr)}}{kr}P(k).
\end{align}
Figure~\ref{xi} shows the two-point correlation functions around the 
baryon acoustic peak at redshifts $z=3$ (top left), $2$ (top right), 
$1$ (bottom left), 
and $0.5$ (bottom right). In each panel, the predictions 
calculated from linear theory (black dotted with error bars), 
1-loop LPT (green dashed) and 2-loop LPT (red solid) are plotted. 
Also, figure~\ref{xi_diff} shows the fractional difference
between the $N$-body and Lagrangian PT up to 1-loop (green) and 2-loop
(red) results respectively, i.e., [$\xi_{N\mbox{\scriptsize -body}}(r)-\xi_{\rm
  PT}(r)]/\xi_{N\mbox{\scriptsize -body}}$.
Note again that 
the error bars of the $N$-body results are rather small, and are visually 
hard to see in figure~\ref{xi}.

Apart from the large scales beyond the location of acoustic peak, where 
the reliability of $N$-body simulations becomes subtle due to the 
limited simulation boxsize~\cite{Taruya09}, 
a good agreement between the $N$-body simulations 
and predictions is found 
for both 1-loop and 2-loop LPT. 
As decreasing redshifts, the acoustic peak structure tends to be erased
and the peak position is shifted to a small scale.
These are purely
the result of
non-linear mode coupling, and the LPT explains this smearing effect quite well \cite{mat08a}.
 On the other hand,  the shift of the baryon acoustic peak 
position is at a sub-percent level.  
Even at redshift $z=0$~\cite{seo10}, it is difficult to 
quantify this effect in our results of $N$-body simulations.
An interesting point is here that the correlation function is 
hardly affected by the 2-loop correction compared to the power spectrum.
This trend has been also reported in other PT treatment (e.g., closure theory
\cite{Taruya09} and RPT \cite{CS08}). 
The difference between $N$-body
simulations and both 1-loop and 2-loop results of LPT 
is small enough around the acoustic peak
($r\sim105\,h^{-1}$\,Mpc), and the accuracy of both LPT predictions 
reaches at several percents level, which is sufficient for ongoing and future 
surveys~\cite{Taruya09}. A closer look at the baryon acoustic peak 
at lower redshifts, however, reveals 
that 2-loop LPT tends to oversmear the peak structure, and the 
agreement with $N$-body simulation looks somewhat degraded. 
Although we have not yet succeeded to identify the origin of this 
discrepancy, the result may suggests either the significance of 
higher-order corrections at lower redshift or the need for delicate
numerical treatment in evaluating the multi-dimensional integration.  
By contrast, a noticeable effect of the 2-loop corrections can be seen
on small scales ($r\lesssim50\,h^{-1}$\,Mpc), where the
nonlinear enhancement of the correlation function becomes prominent at lower
redshifts. Even though the 1-loop LPT tends to deviate from $N$-body 
results, the 2-loop LPT reproduces the $N$-body trends quite well within 1\%
precision.

%===================================================%
\section{Summary \label{sec:summary}}
%===================================================%

In this paper, we present an improved prediction of the nonlinear
 perturbation theory via the Lagrangian picture, 
  originally proposed by ref.~\cite{mat08a}. 
  Based on the previous result \cite{mat08a},  
 we derive a general relation between the power spectrum in SPT and 
 that in LPT for arbitrary loop-order.  
 Using this relation, we explicitly write down the analytic expression of 
 the power spectrum in LPT up to the 2-loop order in both real and redshift 
 spaces.

 Comparing the results in LPT with $N$-body simulations in real space, 
 precisely, we quantitative study the validity range of LPT 
 at various redshifts. For power spectrum, the higher-order corrections  
 enhance the power of high-$k$. 
 Including the 2-loop corrections, 
   the range of validity, where the LPT prediction agrees  
   with $N$-body simulation within $1\%$ precision, is improved 
 by a factor of 1.0 ($z=0.5$), 1.3 ($1$), 1.6 ($2$) and 1.8 ($3$),  
 compared with the 1-loop LPT.
 On the other hand, the two-point correlation functions 
 around the baryon acoustic peak are less sensitive to 
 the higher-order corrections in all redshift range.
 This implies that the 1-loop LPT is sufficient to accurately describe 
 the baryon acoustic peak in correlation functions.

 Finally, we should note remaining issues and future prospects of our work.
 We postponed the evaluation of redshift-space power spectrum and the 
 check of its accuracy. The expression of redshift-space power spectrum 
 involves a difficult term to evaluate, and we need to 
 develop an efficient algorithm to compute it. 
 On the other hand, in practice, it is inevitable to 
 properly take account of the galaxy biasing in the real BAO survey.
 The LPT provides a way to treat the nonlinear 
 galaxy biasing in a self-consistent manner, and the previous 
 study \cite{mat08b} has reported the advantage of LPT 
 based on the 1-loop calculations. The extension to the 2-loop 
 calculations would be presumably a straightforward task, and will be 
 reported elsewhere.

%===================================================%
 \acknowledgments
%===================================================%
 We would like to thank T. Nishimichi for providing us 
 the numerical simulation results and useful comments. 
 This work is supported in part by a Grants-in-Aid for Scientific Research
 from the Japan Society for the Promotion of Science (JSPS) (No.~
 22-2879 for TO, No.~21740168 for AT). TM acknowledges support from 
 the Ministry of Education, Culture, Sports, Science, and 
 Technology, Grant-in-Aid for Scientific Research (C), 21540263, 2009.  
 AT and TM also acknowledge support from the
 Grant-in-Aid for Scientific Research on Priority Areas No.~467
 ``Probing the Dark Energy through an Extremely Wide and Deep Survey
 with Subaru Telescope''. This work is supported in part by JSPS
 (Japan Society for Promotion of Science) Core-to-Core Program
 ``International Research Network for Dark Energy.''
%===================================================%

\end{document}